\documentclass[12pt,showlabels]{iopart}

\newcommand{\be}{\begin{equation}}
\newcommand{\ee}{\end{equation}}
\newcommand{\bea}{\begin{eqnarray}}
\newcommand{\eea}{\end{eqnarray}}
\newcommand{\Q}{{\bf a}}
\newcommand{\B}{{\bf b}}

\newcommand{\RR}{\stackrel{\tiny (3)}{R}}
\newcommand{\RRcal}{\stackrel{\tiny (3)}{\cal R}}

\usepackage{iopams}
\usepackage{psfrag}
\usepackage{latexsym,array,theorem,mathrsfs,subfigure,bm,float}
\usepackage{amsfonts}
\usepackage{amssymb}

\renewcommand{\theequation}{\arabic{section}.\arabic{equation}}

\begin{document}

\title[M.S.Volkov]{Self-accelerating cosmologies and hairy black holes \\
in ghost-free bigravity and massive gravity}

\author{Mikhail S. Volkov}

\address{Laboratoire de Math\'{e}matiques et Physique Th\'{e}orique CNRS-UMR 7350, \\
Universit\'{e} de Tours, Parc de Grandmont, 37200 Tours, FRANCE;\\

\vspace{1 mm}
\vspace{1 mm}

Institut des Hautes Etudes Scientifiques (IHES), 91440 Bures-sur-Yvette, FRANCE}
\ead{volkov@lmpt.univ-tours.fr}
\begin{abstract}
We present a survey of the known cosmological and black hole solutions 
in ghost-free bigravity and massive gravity theories. These
can be divided into three classes. First, there are solutions with 
proportional metrics, which are the same as in General Relativity with 
a cosmological term, which can be positive, negative or zero. 
Secondly, for spherically symmetric systems, there are  
solutions with non-bidiagonal metrics. The g-metric fulfills Einstein
equations with a positive cosmological term and a matter source, while the f-metric
is anti-de Sitter. 
The third class contains solutions with bidiagonal metrics, 
and these can be quite complex. 
The time-dependent solutions describe homogeneous (isotropic 
or anisotropic) cosmologies which show a late-time self-acceleration
or other types of behavior. 
The static solutions describe black holes 
with a massive graviton hair, and also globally regular lumps of energy. 
None  of these 
are asymptotically flat. Including a matter source gives rise to 
asymptotically flat solutions which exhibit the Vainshtein mechanism of recovery 
of General Relativity in a finite region.

\end{abstract}

\maketitle

\section{Introduction}
The recent discovery of a multiparameter class of ghost-free massive gravity 
theories by de Rham, Gabadadze and Tolley (dRGT) \cite{deRham:2010kj}, 
and of its bigravity generalization 
by Hassan and Rosen (HR)
\cite{Hassan:2011zd}, has revived the old idea that 
gravitons can have a small mass $m$ \cite{Fierz:1939ix}
(see \cite{Hinterbichler:2011tt} for a recent review). 
Theories with massive gravitons were, for a long time, considered as 
pathological,  mainly because they exhibit the Boulware-Deser (BD)
ghost -- an unphysical negative norm state in the spectrum \cite{Boulware:1973my}. 
However, there are serious reasons to believe  
that the theories of \cite{deRham:2010kj,Hassan:2011zd} are free of this pathology,
since the unphysical mode is eliminated by an additional constraint,
whose existence is revealed by the canonical analysis \cite{Hassan:2011hr,Golovnev:2011aa,%
Kluson:2012wf,Hassan:2012qv,Hassan:2011ea}, or in the covariant approach
\cite{Deffayet:2012nr}
based on the tetrad formulation \cite{CM}. 
This does not mean that all solutions are 
stable in these theories, since there could be other instabilities,
which should be checked in each particular case. However, since the 
most dangerous BD ghost instability is absent, the theories of 
\cite{deRham:2010kj,Hassan:2011zd}
can be considered as healthy  physical models 
for interpreting the observational data. 

Theories with massive gravitons have been used in order to 
explain the observed acceleration of our universe 
\cite{1538-3881-116-3-1009,0004-637X-517-2-565}. 
This acceleration could be accounted for by introducing a cosmological term in
Einstein equations, however, this would pose the problem of explaining the origin
and value of this term. An alternative possibility is to consider 
{modifications}
of General Relativity (GR), and theories with massive gravitons are {natural}
candidates for this, since the graviton mass can effectively manifest itself
 as a small cosmological term \cite{PhysRevD.66.104025}. 
This motivates 
studying cosmological and other solutions
with massive gravitons. 

Theories with massive gravitons are described by two metrics, $g_{\mu\nu}$
and $f_{\mu\nu}$. In massive gravity theories the f-metric is non-dynamical
and is usually chosen to be flat, although other choices are also possible,
while the dynamical g-metric describes massive gravitons.   
In bigravity theories both metrics are dynamical and describe together 
two gravitons, one of which is massive and the other one is massless.
The theory contains two gravitational couplings, $\kappa_g$ 
and $\kappa_f$,
and in the limit where $\kappa_f$ vanishes, the f-equations decouple from g,
so that the f-metric is determined by its own dynamics. 
The g-equations contain f,
and the g-theory then can be viewed as a massive 
gravity with a fixed background metric f.   
Therefore, massive gravity theory is a special case of bigravity,
and so when  one studies bigravity solutions one finds, in particular, 
all massive gravity ones. 
One should emphasize, however, that in the $\kappa_f\to 0$ limit the f-metric
does not necessarily become flat and the bigravity solutions 
do not always reduce in this limit to solutions of the
standard dRGT massive gravity with flat f.  

The known bigravity solutions can be divided into three types. First,
there are solutions for which the two metrics are proportional in the same coordinate system:
$f_{\mu\nu}=C^2g_{\mu\nu}$.  
This is only possible if $C$ fulfills an algebraic
equation with constant coefficients, and if the matter sources in the two sectors
are fine-tuned to be proportional to each other. The g-equations then reduce 
to the standard Einstein equations with a matter source and a cosmological term
$\Lambda_g\sim m^2$, which can be positive, negative or zero. Therefore,
one obtains in this way all GR solutions, 
in particular, for $C=1$, those of vacuum GR.  
For $\Lambda_g>0$ there are cosmological solutions
which approach the de Sitter space at late times and describe
the cosmic self-acceleration. On the other hand,   
none of the proportional solutions, apart from the trivial one with 
$f_{\mu\nu}=g_{\mu\nu}=\eta_{\mu\nu}$, fulfill equations of the massive gravity 
theory with flat f.

Secondly, imposing spherical symmetry, there are solutions 
described by two metrics which are not bidiagonal, 
although they can be made separately diagonal when expressed 
 in two different frames. 
There is a nontrivial consistency condition relating these two frames,
and as soon as it is fulfilled, 
 the two metrics formally decouple one from the other and 
each of them fulfills its own set of Einstein equations with its own
cosmological and matter terms. 
In some cases, 
the f-metric can be chosen to be anti-de Sitter,
with $\Lambda_f\sim \kappa_f^2$, hence it becomes flat when $\kappa_f\to 0$,
and so the dRGT massive gravity is naturally recovered. 

These solutions  
exist {both} in the dRGT massive gravity and in the HR bigravity theories. 
They describe all known 
massive gravity black holes, in which case the g-metric is Schwarzschild-de Sitter.  
They also describe all known massive gravity cosmologies, for which 
the g-metric is of any (open, closed and flat) 
standard Friedmann-Lem\^{a}itre-Robertson-Walker (FLRW) type. For these solutions 
 the ordinary matter dominates at early times, when the universe
 is small, while later the effective cosmological term $\Lambda_g\sim m^2$
becomes dominant, leading to a self-acceleration. 
Such solutions had been first obtained 
without matter, in which case they describe the pure de Sitter space \cite{Koyama:2011xz,Koyama:2011yg}.
Later the matter term was included,  
\cite{Chamseddine:2011bu,D'Amico:2011jj,Kobayashi:2012fz,Gratia:2012wt},
first only for special values of the theory parameters, 
and then in the general case  
 \cite{Volkov:2012cf,Volkov:2012zb}. If the massive gravity theory were indeed 
the correct theory of gravity, these solutions could describe our
universe. However, perturbations around these FLRW backgrounds  
are expected to be inhomogeneous -- due to the non-diagonal metric components, 
although this effect should be suppressed by the 
smallness of $m$ \cite{D'Amico:2011jj}.
For this reason solutions of this type are sometimes 
called in the literature `inhomogeneous', or solutions with `inhomogeneous 
St\"uckelberg fields'.  

For solutions of the two types discussed above the metrics are the same as in GR
and the graviton mass manifests itself only as an effective cosmological term.  
By contrast, the third type of the known bigravity solutions, where the two metrics are 
bidiagonal in the same frame but not proportional, leads to more 
complex equations of motion  which usually
require a numerical analysis.  The FLRW solutions can be self-accelerating, 
but can also show more complex behaviors 
\cite{Volkov:2011an,vonStrauss:2011mq,Comelli:2011zm,Capozziello:2012re}. 
There are also anisotropic cosmological solutions, and these show 
that the generic state to which the universe
approaches at late times is anisotropic, and the anisotropy energy 
behaves similarly to the energy of a non-relativistic matter \cite{Maeda:2013bha}.     
The static vacuum solutions with bidiagonal metrics can be rather complex and 
describe black holes with a massive graviton hair and also gravitating lumps of energy 
\cite{Volkov:2012wp}. None  of these solutions are asymptotically flat.
The asymptotic flatness is achieved by   
including a regular matter source, which gives rise to 
solutions which exhibit the Vainshtein mechanism of recovery 
of General Relativity in a finite region of space \cite{Vainshtein:1972sx}.

Below, a more detailed description of the currently known 
bigravity and massive gravity solutions is given.


\section{Ghost-free bigravity theory}
\setcounter{equation}{0}
The theory of the ghost-free bigravity \cite {Hassan:2011zd} 
is defined on a four-dimensional spacetime manifold
equipped with two metrics, ${g_{\mu\nu}}$ and 
${f_{\mu\nu}}$, which describe two interacting gravitons, one of which is massive 
and the other one is massless. 
The kinetic term of each metric is chosen to be of the 
standard Einstein-Hilbert form, 
while the interaction between them is 
described by a local potential term ${\cal U}[g,f]$ which does not contain 
derivatives and which is expressed by  
 a scalar function 
of the tensor 
\be                             \label{gam}
\gamma^\mu_{~\nu}=\sqrt{{g}^{\mu\alpha}{f}_{\alpha\nu}}.
\ee
Here ${g}^{\mu\nu}$ is the inverse of ${g}_{\mu\nu}$
and the square root
is understood in the matrix sense, i.e. 
\be                     \label{gamgam}
(\gamma^2)^\mu_{~\nu}\equiv \gamma^\mu_{~\alpha}\gamma^\alpha_{~\nu}={g}^{\mu\alpha}
{f}_{\alpha\nu}.
\ee 
Notice that the matrix square root $\gamma^\mu_{~\nu}$ is well-defined when $g_{\mu\nu}$
and $f_{\mu\nu}$ are close to each other so that ${g}^{\mu\alpha}{f}_{\alpha\nu}$
is close to the unit matrix, in which case $\gamma^\mu_{~\nu}$ is also close to the unit
matrix. However, $\gamma^\mu_{~\nu}$ might cease to be well-defined (and/or real)
for generic g,f.  

Assuming a g-matter and an f-matter 
interacting, respectively, only with ${g}_{\mu\nu}$ 
and with ${f}_{\mu\nu}$,  
the action is (with the metric signature $-+++$) 
\bea                                      \label{1}
&&S[{g},{f},{\rm matter}]
=\frac{1}{2\kappa_g^2}\,\int d^4x\, \sqrt{-{g}}\,R({g})
+\frac{1}{2\kappa_f^2}\,\int d^4x\, \sqrt{-{f}}\, {\cal R}({f})  \nonumber \\
&&-\frac{m^2}{\kappa^2}\int d^4x\, \sqrt{-{g}} \, {\cal U}[{g},{f}] \,
+S^{[\rm m]}_g[{g},{\rm g\mathchar`-matter}]
+S^{[\rm m]}_f[{f},{\rm f\mathchar`-matter}] 
\,, 
\eea
where $R$ and ${\cal R}$ are the Ricci scalars for ${g}_{\mu\nu}$ and 
${f}_{\mu\nu}$, respectively,  $\kappa_g^2=8\pi G$ and $\kappa_f^2=8\pi {\cal G}$
 are the corresponding gravitational couplings, while 
$\kappa^2=\kappa_g^2+\kappa_f^2$ and 
$m$ is the graviton mass.
The interaction 
between the two metrics is given by
\be                             \label{2}
{\cal U}=\sum_{n=0}^4 b_k\,{\cal U}_k(\gamma),
\ee
where $b_k$ are parameters, while  
${\cal U}_k(\gamma)$ are defined 
by the 
relations
\bea                        \label{4}
{\cal U}_0(\gamma)&=&1,  \\
{\cal U}_1(\gamma)&=&
\sum_{A}\lambda_A=[\gamma],\nonumber \\
{\cal U}_2(\gamma)&=&
\sum_{A<B}\lambda_A\lambda_B 
=\frac{1}{2!}([\gamma]^2-[\gamma^2]),\nonumber \\
{\cal U}_3(\gamma)&=&
\sum_{A<B<C}\lambda_A\lambda_B\lambda_C
=
\frac{1}{3!}([\gamma]^3-3[\gamma][\gamma^2]+2[\gamma^3]),\nonumber \\
{\cal U}_4(\gamma)&=&
\lambda_0\lambda_1\lambda_2\lambda_3
=
\frac{1}{4!}([\gamma]^4-6[\gamma]^2[\gamma^2]+8[\gamma][\gamma^3]+3[\gamma^2]^2
-6[\gamma^4])\,. \nonumber 
\eea
Here $\lambda_A$ ($A=0,1,2,3$) are the eigenvalues of $\gamma^\mu_{~\nu}$, 
and, using the hat to denote matrices, one has defined 
$[\gamma]={\rm tr}(\hat{\gamma})\equiv \gamma^\mu_{~\mu}$, 
$[\gamma^k]={\rm tr}(\hat{\gamma}^k)\equiv (\gamma^k)^\mu_{~\mu}$. 
The (real) parameters $b_k$ could be arbitrary, however, if one requires 
flat space to be a solution of the theory, and $m$ to be the Fierz-Pauli 
mass of the graviton \cite{Fierz:1939ix}, 
then the five $b_k$'s are expressed in terms of 
{\it two} free parameters $c_3,c_4$ as follows: 
\bea                \label{bbb}
b_0&=&4c_3+c_4-6,~~b_1=3-3c_3-c_4,~~
b_2=2c_3+c_4-1,~~\nonumber \\
b_3&=&-(c_3+c_4),~~
b_4=c_4.
\eea
The theory (\ref{1}) propagates 7=5+2 degrees of freedom corresponding to 
the polarizations of two gravitons, one massive and one massless. 
Before this theory was discovered 
\cite{Hassan:2011zd}, other bigravity models, sometimes called f-g theories,
had been considered \cite{Isham:1971gm}. 
Such theories propagate 7+1 degrees of freedom, 
the additional one being the BD ghost \cite{Boulware:1973my}.

\subsection{Field equations}
Let us introduce the mixing angle $\eta$ such that
$\kappa_g=\kappa\cos\eta$, $\kappa_f=\kappa\sin\eta$.
Varying the action (\ref{1})  gives the field equations 
\bea                                  \label{Einstein}
G^\mu_\nu&=&m^2\cos^2\eta\, T^{\mu}_{~\nu}
+T^{\rm [m]\mu}_{~~~\nu}\,,~~~~~~~ \\
{\cal G}^\mu_\nu&=&m^2\sin^2\eta\, {\cal T}^{\mu}_{~\nu}+ 
{\cal T}^{\rm [m]\mu}_{~~~\nu}\,, \label{Einstein1}
\eea
where $G^\mu_\nu$ and ${\cal G}^\mu_\nu$ are the  Einstein tensors for $g_{\mu\nu}$
and $f_{\mu\nu}$ and where the two gravitational couplings are included
into the definition of the matter sources $T^{\rm [m]\mu}_{~~~\nu}$
and ${\cal T}^{\rm [m]\mu}_{~~~\nu}$.
The ``graviton" energy-momentum tensors obtained by varying the interaction 
${\cal U}$ are 
\bea                        \label{T}
&&
T^{\mu}_{~\nu}=
\,\tau^\mu_{~\nu}-{\cal U}\,\delta^\mu_\nu,~~~~~
{\cal T}^{\mu}_{~\nu}
=-\frac{\sqrt{-g}}{\sqrt{-f}}\,\tau^\mu_{~\nu}\,,
\eea
where 
\bea                                \label{tau}
\tau^\mu_{~\nu}&=&
\{b_1\,{\cal U}_0+b_2\,{\cal U}_1+b_3\,{\cal U}_2
+b_4\,{\cal U}_3\}\gamma^\mu_{~\nu} \nonumber \\
&-&\{b_2\,{\cal U}_0+b_3\,{\cal U}_1+b_4\,{\cal U}_2\}(\gamma^2)^\mu_{~\nu} \nonumber  \\
&+&\{b_3\,{\cal U}_0+b_4\,{\cal U}_1\}(\gamma^3)^\mu_{~\nu} \nonumber \\
&-&b_4\,{\cal U}_0\,(\gamma^4)^\mu_{~\nu}
\eea
with ${\cal U}_k\equiv {\cal U}_k(\gamma)$. In deriving these expressions 
one uses the 
relations  
\bea
&&
\frac{\delta[\gamma^n]}{\delta g^{\mu\nu}}=\frac{n}{2}\,
g_{\mu\alpha}(\gamma^n)^\alpha_{~\nu}=\frac{n}{2}\,
g_{\nu\alpha}(\gamma^n)^\alpha_{~\mu},\nonumber \\
&&
\frac{\delta[\gamma^n]}{\delta f^{\mu\nu}}=-\frac{n}{2}\,
f_{\mu\alpha}(\gamma^n)^\alpha_{~\nu}=-\frac{n}{2}\,
f_{\nu\alpha}(\gamma^n)^\alpha_{~\mu}
\,,
\eea
which can be obtained by varying the 
definition of $\gamma^\mu_{~\nu}$ 
and using the properties of the trace. 
The matter sources are conserved due to the diffeomorphism invariance 
of the matter terms in the action, 
$\stackrel{(g)}{\nabla}_\mu T^{\rm [m]\mu}_{~~~~~\nu}=0$, 
$\stackrel{(f)}{\nabla}_\mu {\cal T}^{\rm [m]\mu}_{~~~~~\nu}=0$,
where $\stackrel{(g)}{\nabla}$ and $\stackrel{(f)}{\nabla}$
are covariant derivatives with respect to $g_{\mu\nu}$ and 
$f_{\mu\nu}$. 
The Bianchi identities for (\ref{Einstein}) imply that  the graviton
energy-momentum tensor is also conserved, 
$\stackrel{(g)}{\nabla}_\mu\! T^{\mu}_{~\nu}=0$. 
Similarly, the Bianchi identities imply that 
$\stackrel{(f)}{\nabla}_\mu\!{\cal T}^{\mu}_{~\nu}
=0$, but in fact 
this condition is not independent and follows from 
$\stackrel{(g)}{\nabla}_\mu\! T^{\mu}_{~\nu}=0$
 in view of the diffeomorphism invariance of 
the interaction term 
in the action.  

If $\eta\to 0$ and $\sin^2\eta\,{\cal T}^\mu_{~\nu}\to 0$, then 
equations (\ref{Einstein1}) for the f-metric decouple and 
their solution enters the g-equations (\ref{Einstein}) as the fixed reference metric.
The g-equations describe in this case a massive gravity theory.  
For example, if f becomes
flat for $\eta\to 0$, then one recovers the dRGT theory \cite{deRham:2010kj}. 
Therefore, theories of massive gravity are contained in the bigravity. However,
one has to emphasize that the existence of the massive gravity limit is not guaranteed 
\cite{Baccetti:2012bk},
since  $\sin^2\eta\,{\cal T}^\mu_{~\nu}$ does not necessarily vanish when $\eta\to 0$,
and even if it does, the f-metric does not necessarily reduce to something fixed. 
For example, in the absence of the f-matter, 
f-metric becomes Ricci flat in the limit, but this
does not guarantee that it is flat. Therefore, although the massive gravity can be
embedded into bigravity, it is not a limit of the latter but 
a different theory with different properties. As a result, the recent observation that 
the massive gravity theory could be acausal \cite{Deser:2012qx} does not necessarily
apply to the bigravity theory \cite{Hassan:2013pca}.

\section{Proportional backgrounds}
\label{GR}
\setcounter{equation}{0}
The simplest solutions of the bigravity equations are obtained
by assuming the two
metrics to be  proportional, 
\be
f_{\mu\nu}=C^2 g_{\mu\nu}. 
\ee
One obtains  in this case 
$\gamma^\mu_{~\nu}=C\delta^\mu_{~\nu}$ and so 
\bea
\tau^\mu_{~\nu}&=&
(b_1 +3 b_2\,C+3 b_3\,C^2
+ b_4\,C^3)C \delta^\mu_{~\nu}
\,,
\eea
which gives the energy-momentum tensors 
\be
T^{\mu}_{~\nu}=-\Lambda_g(C) \delta^\mu_{~\nu},~~~
{\cal T}^{\mu}_{~\nu}=-\Lambda_f(C) \delta^\mu_{~\nu}\,,
\ee
with 
\begin{eqnarray}                  \label{LAM}
\Lambda_g(C) &=&m^2\cos^2\eta
\left(b_0 +3 b_1\,C+3 b_2\,C^2
+ b_3\,C^3\right)\,,~
\label{Lmbd}  \nonumber 
\\
\Lambda_f(C) &=&m^2\,{\sin^2\eta\over C^3}
\left(b_1 +3 b_2 C\,+3 b_3 C^2\,
+ b_4 C^3\,\right)\,.
\end{eqnarray}
Since  the energy-momentum tensors should be conserved, 
it follows that $C$ is a constant.
As a result, 
one finds two sets of Einstein equations,
\bea      
G_\mu^\nu   +\Lambda_g(C) \delta_\mu^\nu=
T^{\mu}_{~\nu} \,,~~~~
{\cal G}_\mu^\nu    +\Lambda_f(C) \delta_\mu^\nu=
{\cal T}^{\mu}_{~\nu} 
\,.
\eea
Since one has 
${\cal G}_\mu^\nu=  G_\mu^\nu/C^2 $, 
it follows that $\Lambda_f =  \Lambda_g/C^2$,
which gives an algebraic equation for $C$, with $\chi=\tan^2\eta$,
\begin{eqnarray}
\left(b_0 +3 b_1\,C+3 b_2\,C^2
+ b_3\,C^3\right)
={\chi^2\over C}
\left(b_1 +3 b_2 C\,+3 b_3 C^2\,
+ b_4 C^3\,\right)
\label{cosmc}
.
\end{eqnarray}
It follows  also that the matter sources should be fine-tuned such that 
$
{\cal T}^{[{\rm m}]\mu}_{~~~~~\nu} =
T^{[{\rm m}]\mu}_{~~~~~\nu}\,/C^2.
$
Therefore, the independent equations are
the same as in GR,
\bea      
G_\mu^\nu   +\Lambda_g(C) \delta_\mu^\nu=
T^{\mu}_{~\nu} \,.
\eea
If the parameters  $b_k$ are chosen  
according to 
Eq.~(\ref{bbb}), then Eq.(\ref{cosmc}) factorizes,
\bea  \label{LAMC}
0=(C-1)[(c_3+c_4)C^3+(3-5c_3+(\chi-2)c_4)C^2~~~~\nonumber   \\
+((4-3\chi)c_3+(1-2\chi)c_4-6)C+(3c_3+c_4-1)\chi],
\eea
while 
$$
\frac{\Lambda_g}{m^2\cos^2\eta}=(1-C) 
((c_3+C_4)C^2+(3-5c_3-2c_4)C+4c_3+c_4-6). 
$$
Depending on values of $c_3,c_4,\eta$, 
the equation (\ref{LAMC}) can have up to four real roots. 
For example, for $c_3=1$, $c_4=0.3$, $\eta =1$ the roots are
\be                               \label{PP}
C=\{-2.24;\,1;\,0.06;\,3.61\},~~~
\frac{{\Lambda_g}(C)}{m^2}=\{10.12;\,0;\,-0.50;\,-4.50\}.
\ee
As a result, there can be solutions with four different
values of the cosmological constant, which can be positive, negative, or zero. 
If $C=1$ 
then $\Lambda_g=\Lambda_f=0$ and the 
two metrics and the matter sources are identical, 
$g_{\mu\nu}=f_{\mu\nu}$, 
${T}^{[{\rm m}]\mu}_{~~~~~\nu} ={\cal T}^{[{\rm m}]\mu}_{~~~~~\nu}.$
Setting ${T}^{[{\rm m}]\mu}_{~~~~~\nu}=0$, the vacuum GR is recovered, and,
in particular, flat space,  $g_{\mu\nu}=f_{\mu\nu}=\eta_{\mu\nu}$.

Let us consider small fluctuations around flat space, 
$g_{\mu\nu}=\eta_{\mu\nu}+\delta g_{\mu\nu}$, 
 $f_{\mu\nu}=\eta_{\mu\nu}+\delta f_{\mu\nu}$.
Inserting into the general equations (\ref{Einstein}),(\ref{Einstein1}) and 
linearizing with respect to the fluctuations, one finds that the 
linear combinations
\be
h^{\rm (m)}_{\mu\nu}=\cos\eta\,\delta g_{\mu\nu}+\sin\eta\,\delta f_{\mu\nu},~~~
h^{\rm (0)}_{\mu\nu}=\cos\eta\,\delta f_{\mu\nu}-\sin\eta\,\delta g_{\mu\nu}
\ee 
fulfill the Fierz-Pauli equations,
\be
\Box h^{\rm (m)}_{\mu\nu}+\ldots 
=m^2(h^{\rm (m)}_{\mu\nu}-h^{\rm (m)}\eta_{\mu\nu}),~~~~~
\Box h^{\rm (0)}_{\mu\nu}+\ldots =0.
\ee
Therefore, one can identify $h^{\rm (m)}_{\mu\nu}$ and $h^{\rm (0)}_{\mu\nu}$
with the massive and massless graviton fields, respectively. 
It seems that one can explicitly identify in this way 
the massive and massless degrees of freedom
only for fluctuations around  
proportional backgrounds \cite{Hassan:2012wr}. 
Moreover, only for proportional backgrounds 
the null energy condition is fulfilled both in the g and 
f sectors   \cite{Baccetti:2012re}.
Apart from flat space, other solutions with proportional metrics do not admit
the massive gravity limit with flat f-metric.

\section{FLRW cosmologies with non-bidiagonal metrics \label{off}}
\setcounter{equation}{0}
Let us now make a symmetry assumption 
and choose both metrics
to be invariant under spatial SO(3) rotations.
Since the theory is invariant under diffeomorphisms, one can 
choose the spacetime coordinates such that the g-metric is diagonal. 
However, the f-metric will  in general contain an off-diagonal term,
so that  
\bea                             \label{ansatz}
ds_g^2&=&-Q^2dt^2+N^2dr^2+R^2d\Omega^2\,, \nonumber \\
ds_f^2&=&-(aQdt+cNdr)^2+(cQdt-bNdr)^2+u^2R^2d\Omega^2\,.
\eea
Here $Q,N,R,a,b,c,u$ depend on $t,r$ and 
$d\Omega^2=d\vartheta^2+\sin^2\vartheta d\varphi^2$\,. One can take
\be                                  \label{gamma}
\gamma^\mu_{~\nu}=\sqrt{g^{\mu\alpha}f_{\alpha\nu}}=\left(
\begin{array}{cccc}
a & cN/Q & 0 & 0 \\
-cQ/N & b & 0 & 0 \\
0 & 0 & u & 0 \\
0 & 0 & 0 & u
\end{array}
\right)\,,
\ee
whose eigenvalues are 
\be                              \label{lambda}
\lambda_{0,1}=\left.\left.\frac12\right(a+b\pm\sqrt{(a-b)^2-4c^2}\right),
~~~\lambda_2=\lambda_3=u.
\ee
Inserting this  to (\ref{4}) gives 
\bea
{\cal U}_1&=&a+b+2u,~~~~~{\cal U}_2=u(u+2a+2b)+ab+c^2\,, \nonumber \\
{\cal U}_3&=&u\,(au+bu+2ab+2c^2),~~~~~
{\cal U}_4=u^2(ab+c^2). 
\eea
Notice that, although the eigenvalues (\ref{lambda}) can be complex-valued, 
the ${\cal U}_k$'s are always real. 
It is now straightforward to compute the energy-momentum tensors $T^\mu_{~\nu}$
and ${\cal T}^\mu_{~\nu}$ defined by (\ref{T}),(\ref{tau}). In particular,
one finds 
\be                              \label{A20}
T^0_{~r}=\frac{cN}{Q}\,[b_1+2b_2u+b_3u^2].
\ee
It will be assumed in what follows 
that the g-metric is either  static or of the FLRW type,
in which cases there is no radial energy flux and
$T^0_{~r}$ should be zero. Therefore,    
either $c$ should vanish, or the expression in brackets in (\ref{A20}) vanishes.  
The former option will be considered in the next section, 
while presently let us assume that $c\neq 0$ and
\be                         \label{bb}
b_1+2b_2u+b_3u^2=0.
\ee
This yields
\be                                        \label{u}
u=\frac{1}{b_3}\left( 
-b_2\pm\sqrt{b_2^2-b_1 b_3}
\right).
\ee
Notice that $u$ was a priori a function of $t,r$, but now it is restricted to be a constant.  
Using this, one finds that  $T^0_{~0}=T^r_{~r}=-\lambda_g$ and 
${\cal T}^0_{~0}={\cal T}^r_{~r}=-\lambda_f$ where 
\be
\lambda_g=b_0+2b_1u+b_2u^2,~~~~~~\lambda_f=
\frac{b_2+2b_3u+b_4u^2}{u^2}.
\ee
The conditions 
$
\stackrel{(g)}{\nabla}_\rho T^{\rho}_{\lambda}=0\,
$
reduce in this case to the requirement that 
$T^0_{~0}-T^\vartheta_{~\vartheta}$ should vanish. 
On the other hand, one finds 
\be
T^0_{~0}-T^\vartheta_{~\vartheta}=(b_2+b_3 u)
[(u-a)(u-b)+c^2],  \label{cons}
\ee
and since this has to vanish,
either the first or the second factor on the right should be zero. 
The former case was considered in \cite{Chamseddine:2011bu},\cite{Volkov:2011an}
(see also \cite{Kobayashi:2012fz}). 
However, requiring that $b_2+b_3 u=0$ constrains the possible values 
of the theory parameters $b_k$ hence solutions obtained in this way are not general.  
Let us therefore require that \cite{Volkov:2012cf,Volkov:2012zb}
\be                       \label{uuu}
(u-a)(u-b)+c^2=0. 
\ee
Note that in this constraint $u$ is a constant, but $a$, $b$ and $c$ are still 
(unknown) functions of $t,r$. 
In view of this, one has
$T^0_{~0}=T^\vartheta_{~\vartheta}$ and ${\cal T}^0_{~0}
={\cal T}^\vartheta_{~\vartheta}$
hence both energy-momentum tensors are 
proportional to the unit tensor,
$T^\mu_{~\nu}=-\lambda_g\delta^\mu_\nu$ and 
${\cal T}^\mu_{~\nu}=-{\lambda_f}\delta^\mu_\nu$. 
The field equations (\ref{Einstein})  then 
reduce to 
\bea
G^\rho_\lambda+\Lambda_g \delta^\rho_\lambda
&=&T^{{\rm (m)}\,\rho}_{~~~~\lambda} \,,\nonumber \\
{\cal G}^\rho_\lambda+
{\Lambda}_f \delta^\rho_\lambda&=&{\cal T}^{{\rm (m)}\,\rho}_{~~~~\lambda} \,,       
\label{ee2}
\eea
with 
$\Lambda_g=m^2\cos^2\eta\,\lambda$ and 
 ${\Lambda}_f=m^2\sin^2\eta\,{\lambda}_f$. 
As a result, 
the two metrics seemingly decouple one from the other,
and the graviton mass gives rise to two cosmological terms. 
Unlike in the case of proportional metrics, no fine tuning 
between the two matter sources is needed. 
However, one has to remember that solutions
of (\ref{ee2}) 
should fulfill the 
consistency condition (\ref{uuu}). If the parameters $b_k$ are chosen
according to (\ref{bbb}) then $\lambda_g+u^2\lambda_f=-(u-1)^2$, 
therefore, if $\Lambda_g>0$ then 
$\Lambda_f<0$.

Let us consider time-dependent solutions of the FLRW type
and assume the g-matter to be a perfect fluid, 
$T^{{\rm (m)}\,\rho}_{~~~~\lambda}={\rm diag}[-\rho_g(t),P_g(t),P_g(t),P_g(t)]$,
while the f-sector can be chosen to be empty, 
${\cal T}^{{\rm (m)}\,\rho}_{~~~~\lambda}=0$. 
Then the solution of (\ref{ee2}) is chosen to be  
\bea                                \label{gf}
ds_g^2&=&-dt^2+\Q^2(t)\left(\frac{dr^2}{1-kr^2}+
r^2 d\Omega^2\right),~~~~~k=0,\pm 1,\nonumber \\
ds_f^2&=&-\Delta(U)\, dT^2+\frac{dU^2}{\Delta(U)}
+U^2d\Omega^2\,,~~~~~~\Delta=1-\frac{{\Lambda_f}}{3}\,U^2,
\eea
where $\Q$ fulfills the Friedmann equation, 
\be
3\,\frac{\dot{\Q}^2+k}{\Q^2}=\Lambda_g+\rho.
\ee
The g-metric describes 
an expanding FLRW
universe containing the matter $\rho$ and the positive cosmological term $\Lambda_g$.
It can be of any spatial type -- open, closed or flat. 
At early times, when $\Q$ is small, the matter term $\rho$ is dominant,
while later $\Lambda_g$ dominates and the universe enters the 
acceleration phase. 
The f-metric is the anti-de Sitter one expressed in static coordinates. 
One has $\Lambda_f\sim\sin^2\eta\to 0$ 
when $\eta\to 0$, hence  the f-metric becomes flat in this limit. 
Therefore, the 
solutions apply both in the bigravity theory and in dRGT massive gravity. 

\subsection{Imposing the consistency condition}
The described above effective decoupling of the two metrics has been observed
by several authors, in the massive gravity theory   
\cite{Koyama:2011xz,Chamseddine:2011bu,
D'Amico:2011jj,Gratia:2012wt} and also in the bigravity
\cite{Volkov:2011an}. 
However, 
such a decoupling is only possible if the consistency condition 
(\ref{uuu}) is fulfilled, which requires solving a complicated PDE 
(see Eq.(\ref{PDE}) below),
and this has not always been done. 
In   \cite{Koyama:2011xz} the solution was obtained  
for $\rho_g=0$ when the g-metric is the de Sitter one expressed in static 
coordinates, in which case the PDE becomes an ODE. 
A solution for a more general FLRW g-metric  
for $k=0$ and for special values of $b_k$  
was found in \cite{D'Amico:2011jj}, while  the 
general case was considered in 
\cite{Volkov:2012cf,Volkov:2012zb}. 

Let us notice that the two metrics in (\ref{gf}) are expressed in two different
coordinate systems, $t,r$ and $T,U$, whose relation to each other is not yet known.
One has $T=T(t,r)$ and $U=U(t,r)$, so that $dT=\dot{T} dt+T^\prime dr$ and
$dU=\dot{U} dt+U^\prime dr$. Inserting this into the f-metric in (\ref{gf})
and comparing with the f-metric from (\ref{ansatz}), one 
finds the metric coefficients $a,b,c$ in (\ref{ansatz}) expressed in terms 
of  the partial derivatives of $T,U$. 
One also finds that $U=uR=u\,\Q\, r$. 
Inserting the result into 
(\ref{uuu}) gives a non-linear PDE,
\be                       \label{PDE}
\Delta\, \left[\,\Q\,\sqrt{1-kr^2}\,(\dot{U}T^\prime -\dot{T}U^\prime) 
-u^2\Q^2\right]^2=u^2\Q^2A_{+}A_{-},
\ee
with
$A_{\pm}=\Q\,(\Delta\dot{T}\pm\dot{U})+\sqrt{1-kr^2}\,
(U^\prime \pm \Delta T^\prime)$. Since $u$, $\Q$, $U$, $\Delta(U)$ are already known, 
this equation determines $T(t,r)$. 

Let us consider the $\eta\to 0$ limit, when
${\Lambda}_f=0$
and $\Delta=1$. Then one can find 
exact solutions of (\ref{PDE}). If $k=0$ then 
\be                    \label{T1}
T(t,r)=q\int^t \frac{dt}{\dot{\Q}}+\left(\frac{u^2}{4q}+qr^2 \right)\Q\,,
\ee
where $q$ is an integration constant. This solution agrees with the one
obtained in \cite{D'Amico:2011jj} for $c_3=c_4=0$, $u=3/2$. For $k=\pm 1$ one has
\be                   \label{T2}
T(t,r)=\sqrt{q^2+k u^2}\int^t \sqrt{\dot{\Q}^2+k}\, dt+q\Q\sqrt{1-kr^2}.
\ee
If $\Lambda_f\neq 0$ and $\Delta\neq 1$ then exact solutions of (\ref{PDE}) are unknown,
however, at least when ${\Lambda}_f$ is small, solutions  can be constructed 
perturbatively as
$T=T_0+\sum_{n\geq 1}(-{\Lambda_f}/3)^n T_n$. 
Here $T_0$ corresponds to zero order expressions (\ref{T1}),(\ref{T2}),
while the corrections $T_n$ can be obtained by separating the variables with 
the ansatz $T_n=\sum_{m=0}^{n+1} f_m(t) (1-kr^2)^{m/2}$.

This completes the construction, since all field equations and the 
consistency condition are fulfilled.  
Summarizing, one obtains self-accelerating FLRW solutions of all spatial types,
equally valid in the bigravity and massive gravity theories.   
In the latter case
this exhausts all possible homogeneous and isotropic  cosmologies. 
Somewhat confusingly, these solutions are sometimes called in the 
literature inhomogeneous, because the two metrics are non-bidiagonal and 
$T,U$, which
play the role of St\"uckelberg scalars in the massive gravity limit, 
are inhomogeneous,
as they both depend on $r$. 
It is then expected that fluctuations around the 
FLRW backgrounds should show an inhomogeneous spectrum,  
 although this effect will be 
proportional to $m^2$ and so will be small  \cite{D'Amico:2011jj}. 

In the spatially open case, $k=-1$, choosing in (\ref{T2}) $q=u$, 
 yields $T=u\Q\sqrt{1+r^2}$. Inserting this to (\ref{gf}), 
together with $U=u\Q r$, 
 gives the flat f-metric which turns out to be diagonal {\it both} 
in the $T,U$ and $t,r$ coordinates,
\be                                \label{gf1}
ds_f^2=-dT^2+dU^2+U^2d\Omega^2=u^2\Q^2\left(-\frac{\dot{\Q}^2}{\Q^2}\,dt^2+\frac{dr^2}{1+r^2}+
r^2 d\Omega^2\right),
\ee
so that the g and f metrics happen to be bidiagonal in this case. 
This particular solution was  found in \cite{Gumrukcuoglu:2011ew}.
It is unclear if it should be physically distinguished
when compared to the other solutions  in (\ref{T1}),(\ref{T2})
for generic $k,q$. The analysis of  \cite{DeFelice:2012mx} shows that this 
solution is unstable at the nonlinear level, while the case of generic $k,q$ 
has not yet been studied.

\section{FLRW cosmologies with bidiagonal metrics}
\setcounter{equation}{0}
All solutions considered above are described by the same metrics as in GR,
and the graviton mass manifests itself only as the effective cosmological
term(s). More general solutions are obtained assuming that 
both metrics are simultaneously diagonal 
\cite{Volkov:2011an,vonStrauss:2011mq,Comelli:2011zm,Visser,Capozziello:2012re}, 
\bea                                \label{ggg}
ds_g^2&=&-dt^2+\Q(t)^2\left(\frac{dr^2}{1-kr^2}+
r^2 d\Omega^2\right),~~~\nonumber \\
ds_f^2&=&-{\cal A}(t)^2dt^2+\B(t)^2\left(\frac{dr^2}{1-kr^2}+
r^2 d\Omega^2\right).
\eea
The $G^0_0$ and ${\cal G}^0_0$ equations (\ref{Einstein}),(\ref{Einstein1}) then read
\bea                        \label{OW}
\frac{\dot{\Q}^2}{\Q^2}=
\frac{\Lambda_g(\xi)+\rho_g}{3}-\frac{k}{\Q^2},~~~~~
\frac{1}{{\cal A}^2}\frac{\dot{\B}^2}{\B^2}=
\frac{\Lambda_f(\xi)+\rho_f}{3}-\frac{k}{\B^2}\,
\,,~~~~\label{C2a} 
\eea
where $\xi=\B/\Q$ and where $\Lambda_g,\Lambda_f$ are defined in (\ref{LAM}).
The 
$\stackrel{(g)}{\nabla}_\mu\! T^{\mu}_{~\nu}=0$ conditions reduce to
\be                         \label{factors}
\left[\dot{\B}-{\cal A}\dot{\Q}\right]
\left(b_1 +2b_2\xi+
b_3 \xi^2\right)=0. 
\ee

\subsection{Generic solutions}
Let us assume that the first factor in (\ref{factors}) vanishes, 
$\dot{\B}-{\cal A}\dot{\Q}=0$. 
Comparing this to
(\ref{OW}) gives the algebraic relation 
$
\Lambda_g(\xi)+\rho_g=
\xi^2(\Lambda_f(\xi)+\rho_f),
$ 
or, setting for simplicity $\rho_f=0$, 
\be                           \label{base0}
\frac{b_3}{3}\,\xi^4+(b_2-\frac{\chi}{3} b_4)\,\xi^3+(b_1-\xi b_3)\,\xi^2
+(\frac{b_0}{3}-\chi b_2+\frac{\rho_g}{3\cos^2\eta})\,\xi=\frac{\xi b_1}{3}\,,
\ee
with $\chi=\tan^2\eta$. 
Assuming the equation of state $P_g=w_g\rho_g$ one has
$\rho_g=\rho_g^0\,\Q^{-3(1+w_g)}$, and so that Eq.(\ref{base0}) gives
an algebraic relation between $\Q$ and $\xi$, 
whose solution is $\xi(\Q)$. 
Inserting $\xi(\Q)$ into the first equation in (\ref{OW}) gives the
 Friedmann equation, 
\be                        \label{QQQ}
\dot{\Q}^2+V(\Q)=-k\,,~~~~~~~V(\Q)=-\frac{\Q^2}{3}(\Lambda_g(\xi)+\rho_g),
\ee
where $\xi=\xi(\Q)$ and $\rho_g=\rho_g(\Q)$, hence  $\Q$
is confined to the region where $V(\Q)\leq -k$.

\begin{figure}[th]
\hbox to \linewidth{ \hss	
	\psfrag{x}{$\ln(\Q)$}
	\psfrag{xi1}{$\xi_1$}
        \psfrag{xi2}{$\xi_2$}
	\psfrag{V1}{$V_1$}
	\psfrag{V2}{$V_2$}
	\resizebox{7.8cm}{5cm}{\includegraphics{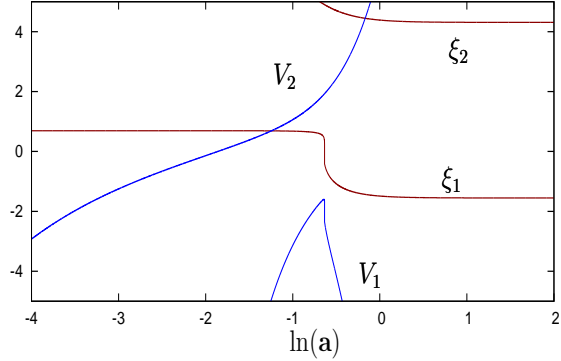}}
\hspace{1mm}
	
\hspace{1mm}
\hss}
\caption{{\protect\small 
The two solutions $\xi_1$, $\xi_2$ of (\ref{base0}) and the corresponding effective 
potentials ${\rm V}(\Q)$ for the same parameter values as in (\ref{PP}). 
One has $V_1<-1$, hence for any $k=0,\pm 1$
solutions of (\ref{QQQ}) travel over the $V_1$ potential barrier 
and enter the accelerated phase for large $\Q$. 
Since $\Lambda_g(\xi_2)<0$, $V_2\to +\infty$ for large $\Q$
and solutions of (\ref{QQQ})
bounce from the potential barrier and recollapse.    }}%
\label{fig1}
\end{figure}

Depending on choice of the solution of 
(\ref{base0}), the shape of the potential $V(\Q)$ can be different. 
The simplest solutions are obtained for $b_1=b_3=0$ 
\cite{vonStrauss:2011mq}. 
Making the parameter choice (\ref{bbb}), the
 self-accelerating solutions are found only if $b_1=3-3c_3-c_4<0$ 
\cite{Volkov:2011an}. 
In this case Eq.(\ref{base0}) has only two real roots $\xi_1(\Q)$ and 
$\xi_2(\Q)$  which exist for any positive 
$\Q$, $\Lambda_g(\xi_1)$ approaching a positive
value at large $\Q$ while $\Lambda_g(\xi_2)$ being negative. 
This corresponds, respectively, to the accelerating and recollapsing
solutions (see Fig.\ref{fig1}).  
For $b_1>0$ there could be more complex 
solutions for which 
the universe size oscillates around a finite value \cite{Volkov:2011an}. 
When Eq.(\ref{QQQ}) is solved, the f-metric
coefficients are $\B=\Q\,\xi(\Q)$ and 
${\cal A}=\dot{\B}/\dot{\Q}$. 
The perturbative analysis reveals that the self-accelerating  
solutions are stable \cite{Comelli:2012db,Berg:2012kn,Khosravi:2012rk},
hence they  can be used for interpreting observational data \cite{Akrami:2013pna}. 
Cosmological solutions with bidiagonal metrics have also been studied 
in the generalized $F(R)$-type bigravity theory \cite{Odintsov}.

\subsection{Special solutions \label{special}}
Let us now assume that the second  factor in (\ref{factors}) vanishes, 
$
{b_1 +2b_2\xi+
b_3 \xi^2}=0.
$
Notice that this is the same equation as in (\ref{bb}). It follows that 
$\xi=\B/\Q$ is constant, therefore $\dot{\Q}/\Q=\dot{\B}/\B$.
Equations then again reduce to (\ref{QQQ}),
where $\Lambda_g$ is now constant.
Combining the two equations 
(\ref{OW}) gives 
\be                        \label{AAA}
{\cal A}^2=\frac{\xi^2\dot{\Q}^2}
{\xi^2\Q^2(\Lambda_f+\rho_f)/3-k},
\ee
and this expression can be positive 
only in special cases. 
In the massive gravity limit, where $\rho_f=\eta=\Lambda_f=0$,  
${\cal A}^2$ is positive only for $k=-1$, which reproduces the solution
(\ref{gf1}), and ${\cal A}$ does not exist for $k=0,1$, hence 
there are no FLRW cosmologies with bidiagonal metrics in these cases 
\cite{D'Amico:2011jj}.

\section{Anisotropic cosmologies}
\setcounter{equation}{0}
Solutions considered in the 
previous section can be generalized to the anisotropic case 
by choosing the two metrics to be bidiagonal and of the same Bianchi type
\cite{Maeda:2013bha},
\bea                                                    \label{BianA}
ds_g^2=-\alpha^2dt^2+h_{ab}\,\omega^a\otimes \omega^b,~~~ 
ds_f^2=-{\cal A}^2dt^2+{\cal H}_{ab}\,\omega^a\otimes\omega^b\,.
\eea
Here $\alpha,h_{ab},{\cal A},{\cal H}_{ab}$ depend on $t$, and the vectors
$e_a$ dual to the one-forms $\omega^a$ generate a three-parameter translation
group acting on the 3-space,
$[e_a,e_b]=C^c_{~ab}e_c$. The structure coefficients are 
$C^c_{~ab}=n^{cd}\epsilon_{dab}$ with
$n^{ab}={\rm diag}[n^{(1)},n^{(2)},n^{(3)}]$, where 
$n^{(a)}$ assume the values 
shown in Table 1.

This choice of the structure coefficients corresponds to the 
Bianchi types of class A,
in which case one can choose 
$
h_{ab}={\rm diag}[\alpha_1^{\,2},\alpha_2^{\,2},\alpha_3^{\,2}]$ and 
$
{\cal H}_{ab}=
{\rm diag}[{\cal A}_1^{\,2},{\cal A}_2^{\,2},{\cal A}_3^{\,2}],
$
and this guarantees that $G^0_r={\cal G}^0_r=0$, so that  
radial energy fluxes are absent. Choosing 
\bea
[\alpha_1,\alpha_2,\alpha_3]&=&e^{\Omega}[
e^{\beta_{+}+\sqrt{3}\beta_{-}} , e^{\beta_{+}-\sqrt{3}\beta_{-}},
 e^{-2\beta_{+}} ], \nonumber \\
\left[
{\cal A}_1,
{\cal A}_2,
{\cal A}_3
\right]&=&
e^{\cal W}[
e^{{\cal B}_{+}+\sqrt{3}{\cal B}_{-}},
e^{ {\cal B}_{+}-\sqrt{3}{\cal B}_{-}} ,
e^{-2{\cal B}_{+}}],
\eea
the spatial 3-curvature of the g-metric is
\be
\RR=
\frac{2n^{(1)}n^{(2)}}{\alpha_3^2}
-\frac{1}{2\alpha_1^2\alpha_2^2\alpha_3^2}\left(
n^{(1)}\alpha_1^2
+n^{(2)}\alpha_2^2
-n^{(3)}\alpha_3^2\right)^2,
\ee
while the 3-curvature $\RRcal$ of the f-metric is obtained from this by 
replacing $\alpha_a\to {\cal A}_a$ and $\Omega\to{\cal W}$. 
\begin{center}
\begin{table}~~~~~~~~~~~~~~~~~~~
\begin{tabular}{|c|c|c|c|c|c|c|}
\hline
    & ~I~~ & ~II~ & ~VI$_0$ & VII$_0$ & VIII & ~IX~ \\
\hline
$n^{(1)}$ & $0$ & $1$ &  $1$ & $1$ & $1$ & $1$ \\
$n^{(2)}$ & $0$ & $0$ & $-1$ & $1$ & $1$ & $1$ \\
$n^{(3)}$ & $0$ & $0$ &  $0$ & $0$ & $-1$ & $1$ \\
\hline
\end{tabular}
\caption{Values of $n^{(a)}$ for the Bianchi class A types}
\end{table}
\end{center}
The potential (\ref{2}) is ${\cal U}\sqrt{-g}=\alpha U_g+{\cal A}U_f$ where 
\bea
U_g&=&
b_0 e^{3\Omega}+b_3 e^{3{\cal W}}
+b_1e^{{\cal W}+2\Omega}
\left(e^{-2({\cal B}_+-\beta_+)}+
2e^{{\cal B}_+-\beta_+}\cosh[\sqrt{3}({\cal B}_--\beta_-)]
\right)
\nonumber \\
&+&b_2 e^{2{\cal W}+\Omega}\left(e^{2({\cal B}_+-\beta_+)}+
2e^{-({\cal B}_+-\beta_+)}\cosh[\sqrt{3}({\cal B}_--\beta_-)]\right),
\eea
while $U_f$ is obtained from this by replacing 
$b_k\to b_{k+1}$. 
The equations for the g-metric read 
\bea
{ \left(\frac{\dot{\Omega}}{\alpha}\right)^2}&=&{
\left(\frac{\dot{\beta}_{+}}{\alpha}\right)^2
+\left(\frac{\dot{\beta}_{-}}{\alpha}\right)^2
+\frac16\left[2\cos^2\eta\,  e^{-3\Omega}U_g
-\RR
+2{\rho_g}\right]}\,,\label{C1} \nonumber \\
\left(e^{3\Omega}\,\frac{\dot{\Omega}}{\alpha}\right)^{\centerdot}
&=&{1\over 6}\left[
\cos^2\eta  \left(\frac{\partial U}{\partial \Omega}+3\alpha\,{\cal U}_g\right)
-2\alpha \,e^{3\Omega}\RR
+3 \alpha \,e^{3\Omega}(\rho_g-P_g)\right]\,, \label{e1}\nonumber\\
\left(e^{3\Omega}\,\frac{\dot{\beta}_{\pm}}{\alpha}\right)^{\centerdot}
&=&-{1\over 12}
\frac{\partial}{\partial{\beta_\pm}} \left(2 \cos^2\eta\,  U-{\alpha}\,
e^{3\Omega}\RR\right),     
\eea
while the equations for the f-metric are obtained from these by replacing 
$\Omega\to{\cal W}$, $\beta_\pm\to{\cal W}_\pm$, $\RR\to\RRcal$, 
${\cal U}_g\to {\cal U}_f$ and $\cos\eta\to\sin\eta$. The conditions 
$\stackrel{(g)}{\nabla}_\mu\! T^{\mu}_{~\nu}=0$ reduce to 
$$
\alpha\,\left(\dot{W}\frac{\partial}{\partial{\cal {\cal W}}}
+\dot{\cal B}_{+}\frac{\partial}{\partial{\cal {\cal B}_{+}}}
+\dot{\cal B}_{-}\frac{\partial}{\partial{\cal {\cal B}_{-}}}\right)U_g=
{\cal A}\,\left(\dot{\Omega}\frac{\partial}{\partial\Omega}
+\dot{\beta}_{+}\frac{\partial}{\partial{\beta_{+}}}
+\dot{\beta}_{-}\frac{\partial}{\partial{\beta_{-}}}
\right){\cal U}_f\,.
$$
All quantities in these equations are dimensionless,
assuming $1/m$ to be the length scale,
while the energy density is measured
in units of $m^2 M_{\rm pl}^2$.

For the Bianchi types I or IX one can 
set anisotropies to zero, $\beta_\pm={\cal B}_\pm=0$,
and then, with $\Q=2e^\Omega$ and $\B=2e^{\cal W}$, the above equations 
reduce to  the FLRW equations (\ref{OW}),(\ref{factors}) for $k=0,1$
(the spatially open $k=-1$ FLRW case is contained in the Bianchi V type, 
which does not belong to the class A).  


In the Bianchi I case it is consistent to set
$
\beta_\pm={\cal B}_\pm\,
$
and then the two metrics are proportional as described in section \ref{GR},
$f_{\mu\nu}=C^2 g_{\mu\nu}$, if only the matter sources are fine-tuned such that 
$\rho_f=\rho_g/C^2$. One has 
\be                                  \label{BNI}
ds_g^2=-dt^2+e^{2\Omega}\left(e^{2\beta_{+}+\sqrt{3}\beta_{-}}dx_1^2
+e^{2\beta_{+}-\sqrt{3}\beta_{-}}dx_2^2
+e^{-4\beta_{+}}dx_3^2\right),
\ee
where $\Omega,\beta_\pm$ fulfill (assuming that $\alpha=1$) 
\bea                              \label{BianI}
\dot{\Omega}^2&=&
(\sigma_{+}^2+\sigma_{-}^2)\, e^{-6\Omega}
+\frac{1}{3}\left({\Lambda_g(C)}+\rho_g \right), \nonumber \\
\dot{\beta}_\pm&=&\sigma_\pm e^{-3\Omega}\,,
\eea
with constant $\sigma_\pm$. Here $C$, $\Lambda_g(C)\equiv 3H^2$ are defined 
by (\ref{LAM}),(\ref{cosmc}). At late times solutions of Eqs.(\ref{BianI}) 
 approach 
configurations with constant anisotropies, 
$\Omega=Ht+O(e^{-3Ht})$, 
$\beta_\pm=\beta_\pm(\infty)+O(e^{-3Ht})$.  


It turns out that solutions for all other Bianchi types and for 
non fine-tuned sources also evolve into  
configurations with equal and nonvanishing 
anisotropies,  $\beta_\pm={\cal B}_\pm$.
This can be seen by applying a numerical procedure 
\cite{Maeda:2013bha}, whose input configuration  
at the initial time moment $t=0$ is an anisotropic deformation 
of a finite size FLRW universe. The initial value of $\Omega$ is zero, hence  
the initial universe size $e^\Omega\sim 1$ (in {1/m} units), while  
the initial anisotropies are $\beta_\pm$, ${\cal B}_\pm$, $\dot{\beta}_\pm$, 
$\dot{\cal B}_\pm$ $\sim 10^{-2}$.  
The initial values of $\dot{\Omega}$, ${\cal W}$, $\dot{\cal W}$ are determined
by resolving the first order constraints contained in the field
equations. It is assumed that 
 the f-sector is empty, $\rho_f=0$,
whereas the g-sector contains a radiation and a non-relativistic matter,
$\rho_g=0.25\times  e^{-4\Omega}+0.25\times e^{-3\Omega}$, so that 
the dimensionful energy ${m^2}M_{\rm pl}^2\,\rho_g\sim 10^{-10}({\rm eV})^4$,
assuming that ${m}\sim 10^{-33}{\rm eV}$. 

 The numerical extension to $t>0$ of the $t=0$ initial data reveals that,
for all Bianchi types, the universe approaches the state with proportional metrics
in which the expansion rate $\dot{\Omega}$ is constant, $H$, and the anisotropies
$\beta_\pm={\cal B}_\pm$ are also constant.  
For the Bianchi I solutions the constant anisotropies can be absorbed 
by redefining the spatial coordinates, but not for other 
Bianchi types. This means that the universe generically runs into anisotropic
states at late times.
\begin{figure}[h]
\hbox to \linewidth{ \hss
	

	\resizebox{7.8cm}{5cm}{\includegraphics{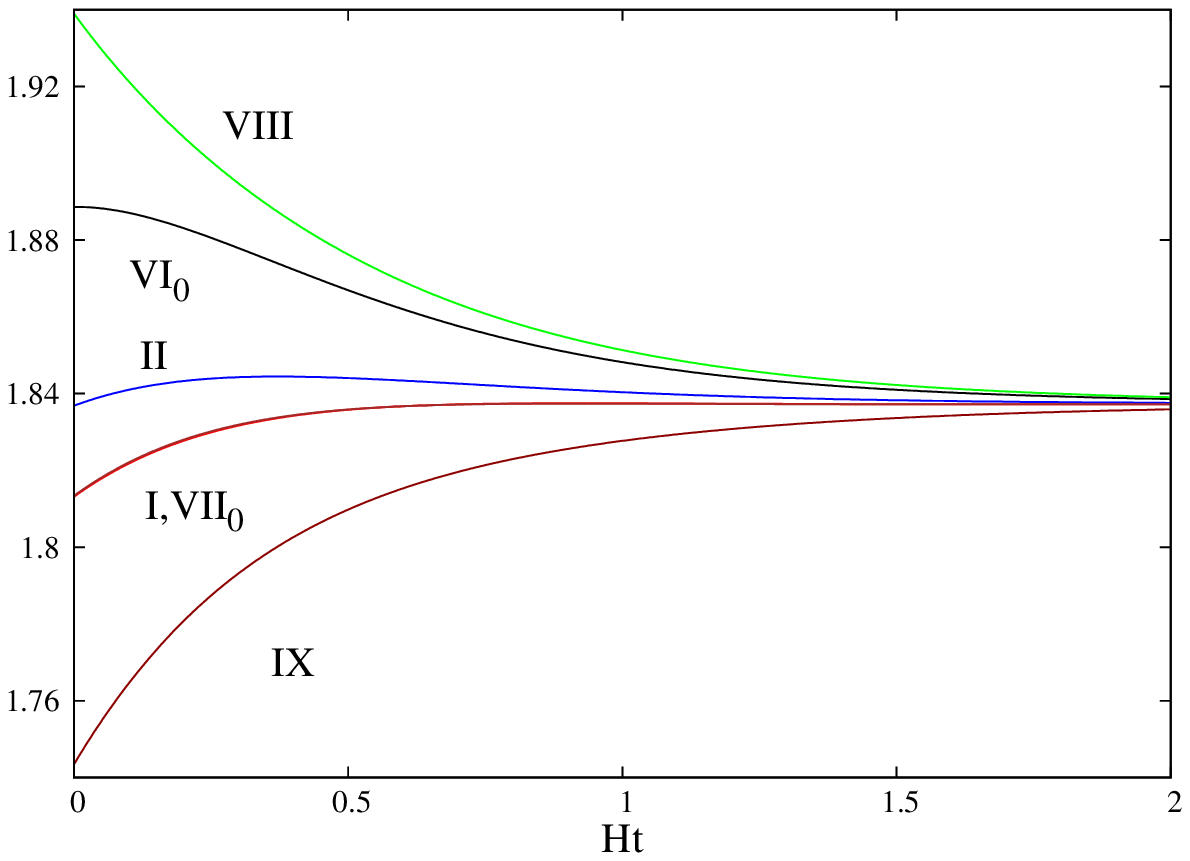}}
\hspace{1mm}
	\resizebox{7.8cm}{5cm}{\includegraphics{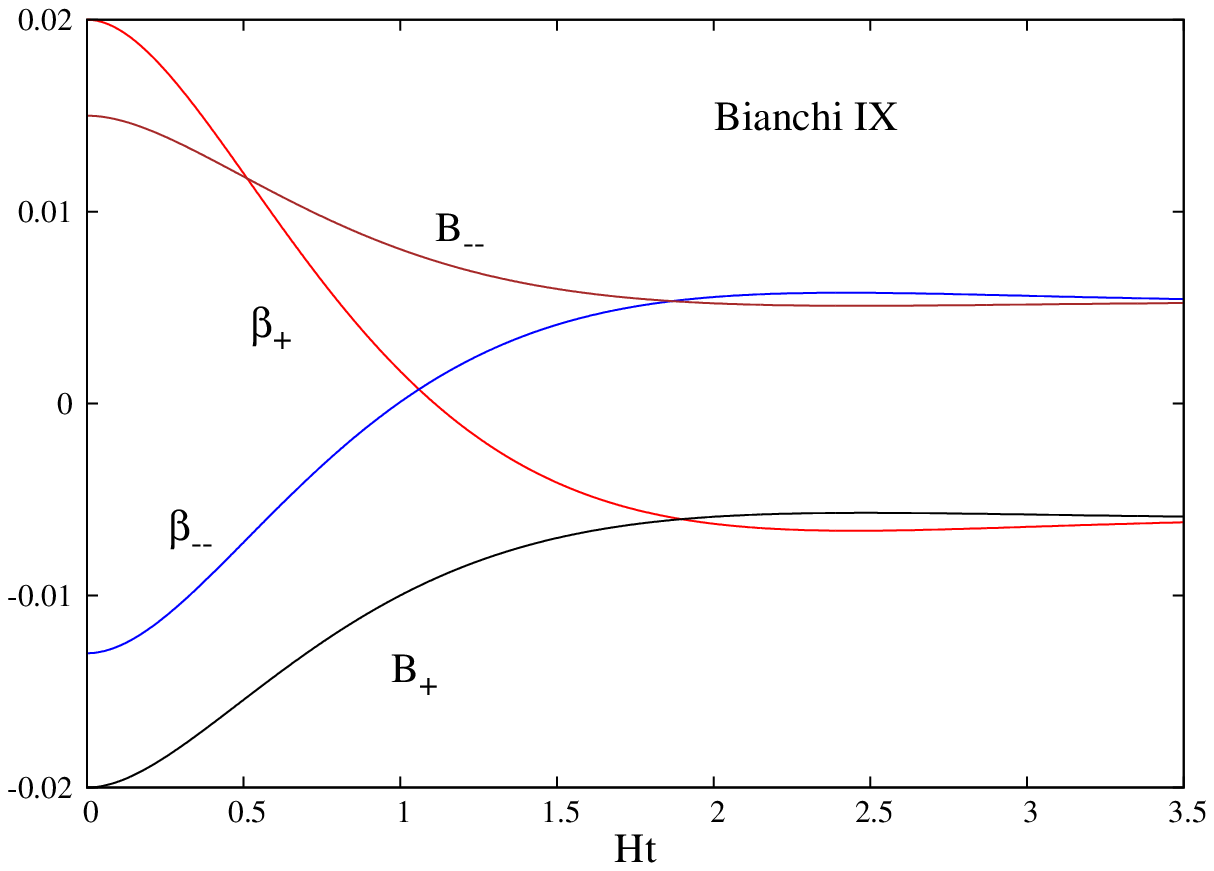}}
	
\hspace{1mm}
\hss}
\caption{{\protect\small The Hubble parameter $\dot{\Omega}$
for all Bianchi types of class A (left) and the anisotropy amplitude for 
the Bianchi IX solution (right). 
}}%
\label{Fig2}
\end{figure}
These properties of the solutions are illustrated in Fig.\ref{Fig2}.  
The relative shear contribution to the total energy density 
$\Sigma=\sqrt{\dot{\beta}^2_{+}+\dot{\beta}^2_{-}}/\dot{\Omega}$ 
tends to zero at late times, however, 
if only just one or two Hubble times have elapsed since the beginning of the 
current phase of the universe acceleration, $\Sigma$ should not necessarily be small. 
It turns out that at late times the anisotropies oscillate around their 
asymptotic values. Linearizing the field equations with respect to small
deviations form the proportional background $f_{\mu\nu}=C^2g_{\mu\nu}$ to which the solutions approach, one finds that   
\be
\dot{\beta}_\pm\sim \dot{\cal B}_\pm\sim e^{-3Ht/2}\cos(H\omega t),
\ee 
where 
$\omega$ can be expressed in terms of $C,b_k,{\eta}$ \cite{Maeda:2013bha}.
It follows that 
the shear energy is {
${
\dot{\beta}_{+}^2+\dot{\beta}_{-}^2\sim e^{-3\Omega}\sim 1/{\bf a}^3\,}
$},
while  in GR it decreases as  $\sim 1/\Q^6$. The fact that the shear energy 
in the bigravity theory 
shows the same fall-off rate as a cold dark matter suggests that the 
latter could in fact be an anisotropy effect, although it is unclear if this 
interpretation can also explain the dark matter clustering.  
It is interesting that the interpretation of the cold dark matter
as an effect of the graviton mass has already been discussed,
although  within a different theory 
with massive gravitons \cite{Dubovsky}.

\begin{figure}[h]
\hbox to \linewidth{ \hss
	

	\resizebox{7.8cm}{5cm}{\includegraphics{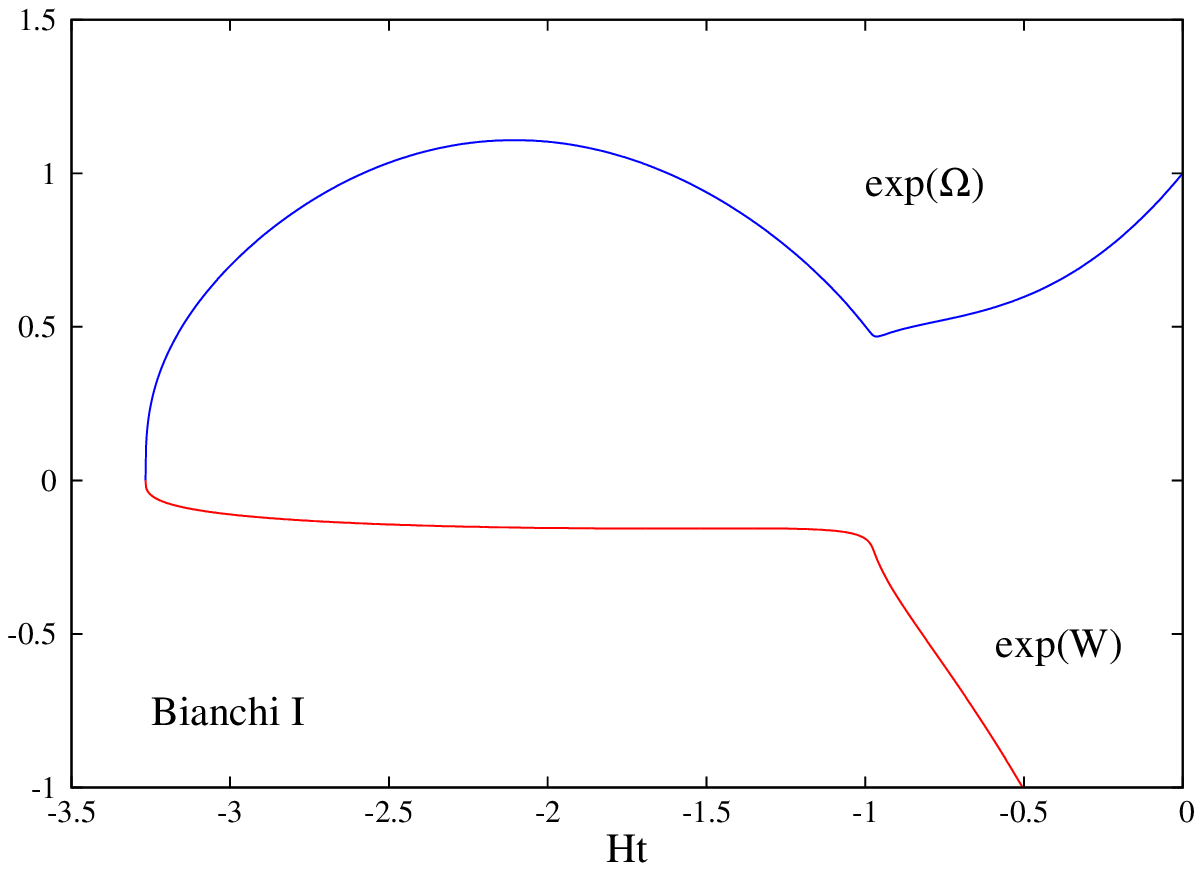}}
\hspace{1mm}
	\resizebox{7.8cm}{5cm}{\includegraphics{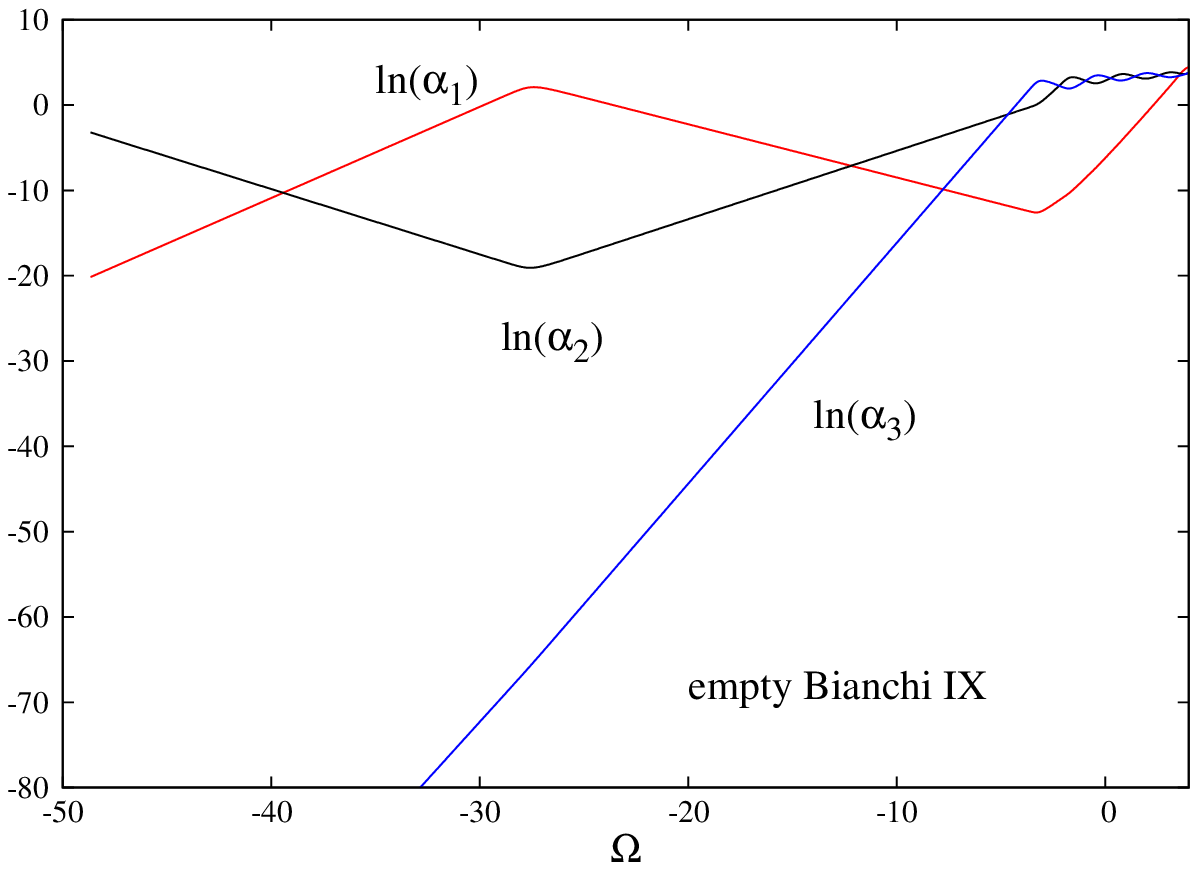}}
	
\hspace{1mm}
\hss}
\caption{{\protect\small Near singularity behavior for the Bianchi I (left)
and Bianchi IX (right) solutions. 
}}%
\label{Fig4}
\end{figure}
When continued to the past, the solutions hit a singularity where 
both $e^\Omega$ and $e^{\cal W}$ vanish (see Fig.\ref{Fig4}). 
The Bianchi IX solutions, when approaching 
singularity,  show the typical billiard behavior 
characterized by a sequence of Kasner-type periods \cite{BKL}. During 
each period one has $\alpha_p\propto t^{p_a}$ where 
$p_1+p_2+p_3=p_1^2+p_2^2+p_3^2=1$,
such that two of the three $\alpha_a$'s grow in time whereas  the third one decreases,
until the next period starts when one of the growing amplitudes 
becomes decreasing (see Fig.\ref{Fig4}).

\section{Black holes, lumps, stars and the Vainshtein mechanism}
\setcounter{equation}{0}

Historically, the first bigravity black holes 
were obtained in the generic f-g bigravity theory \cite{Isham:1971gm}
for non-bidiagonal metrics 
\cite{Isham:1977rj}. Their counterparts in the  ghost-free 
bigravity under consideration are described by the ansatz (\ref{ansatz}).
The field equations then reduce to (\ref{ee2}) (without matter terms), 
and the solution is the Schwarzschild-de Sitter metric,
\be                          \label{SdS}
ds_g^2=-Ddt^2+\frac{dr^2}{D}+r^2d\Omega^2,~~~~D=1-\frac{2M}{r}-\frac{\Lambda_g r^2}{3},
\ee
with $\Lambda_g$ defined after Eq.(\ref{ee2}). 
The f-metric is still expressed   in terms of $T,U$ by (\ref{gf}), and 
to fulfill the 
consistency condition (\ref{uuu}) the procedure is the same as 
in Section~\ref{off}, which gives $U=ur$ and 
$T=ut-u\int \frac{D-\Delta}{D\Delta}\,dr$ with $u$ from (\ref{u})
\cite{Volkov:2012wp}. Since the f-metric becomes flat for $\eta\to 0$,
this solution describes  black holes also in the dRGT massive gravity 
\cite{Koyama:2011xz,Koyama:2011yg} 
(see also \cite{Berezhiani:2011mt}). 

These solutions and their generalizations 
for a nonzero electric charge \cite{Sarid}
exhaust all known black holes in the massive gravity theory.
In particular, there are no asymptotically flat black holes in this case 
(see \cite{Gruz} for a recent discussion). 
However, in the bigravity one finds more solutions  when 
the metrics are bidiagonal \cite{Volkov:2012wp}, 
\bea
ds_g^2&=&Q^2dt^2-\frac{dr^2}{N^2}-r^2d\Omega^2,~~~~
ds_f^2=A^2 dt^2-\frac{U^{\prime 2}}{Y^2} dr^2-U^2d\Omega^2\,.
\eea
Here 
$Q,N,Y ,U,a$ are 5 functions of $r$ which fulfill  
the equations 
\bea                              \label{eqs}
G^0_0&=&m^2{\cos^2\eta}\, T^0_0,~~~~~~~~
G^r_r=m^2\,{\cos^2\eta}\, T^r_r,~~\nonumber \\
{\cal G}^0_0&=&m^2\,{\sin^2\eta}\, {\cal T}^0_0,~~~~~~~
{\cal G}^r_r=m^2\,{\sin^2\eta}\, {\cal T}^r_r,\nonumber \\
{T^r_r}^\prime &+&\frac{Q^\prime}{Q}\,(T^r_r-T^0_0)
+\frac{2}{r}(T^\vartheta_\vartheta-T^r_r)=0.   
\eea 
The simplest solutions are again obtained for the proportional
metrics, $f_{\mu\nu}=C^2g_{\mu\nu}$, with $g_{\mu\nu}$ given by 
(\ref{SdS}) where $\Lambda_g=\Lambda_g(C)$ is defined by 
(\ref{Lmbd}),(\ref{cosmc}). Since $\Lambda_g$ can 
be positive, negative, or zero, there are the
Schwarzschild (S), Schwarzschild-de Sitter (SdS), and 
Schwarzschild-anti-de Sitter (SAdS) black holes.  
Let us call them background black holes. More general 
solutions are obtained by numerically integrating Eqs.(\ref{eqs}). 

In doing this, it is assumed that the g-metric has a regular 
event horizon at $r=r_h$, where $T^\mu_\nu$  and the curvature are finite. 
A detailed analysis shows \cite{Volkov:2012wp}
that the horizon should be common 
for both metrics, therefore amplitudes $Q^2$, $N^2$, $A^2$, $Y^2$ should 
have a simple zero, while $U,U^\prime$ should be non-zero at $r=r_h$. Next, one finds 
that the boundary conditions at the horizon comprise a one-parameter family
labeled by $u=U(r_h)/r_h$, the ratio of the event horizon radius 
measured by the f-metric to that measured by the g-metric. Finally,
it turns out that the horizon surface gravities and temperatures 
determined with respect to both metrics are the same \cite{Deffayet:2011rh}. 

Choosing a value of $u$ and integrating numerically the 
equations from $r=r_h$ towards large $r$, the result is as follows \cite{Volkov:2012wp}. 
If $u=C$ where $C$ fulfills Eq.(\ref{cosmc}) then the solution 
is one of the background black holes described above. If $u=C+\delta u$
with a small $\delta u$, then one can expect the solution to be the   
background black hole slightly deformed by a massive graviton `hair' localized
in the horizon vicinity. 
\begin{figure}[th]
\hbox to \linewidth{ \hss

	\resizebox{8cm}{5cm}{\includegraphics{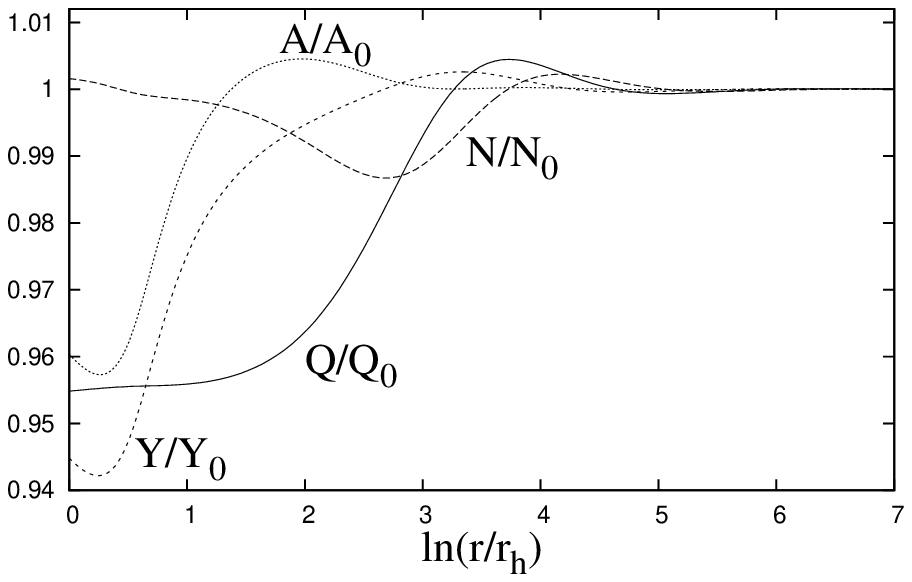}}
\hspace{1mm}
	\psfrag{y}{$U^\prime$}
	\resizebox{8cm}{5cm}{\includegraphics{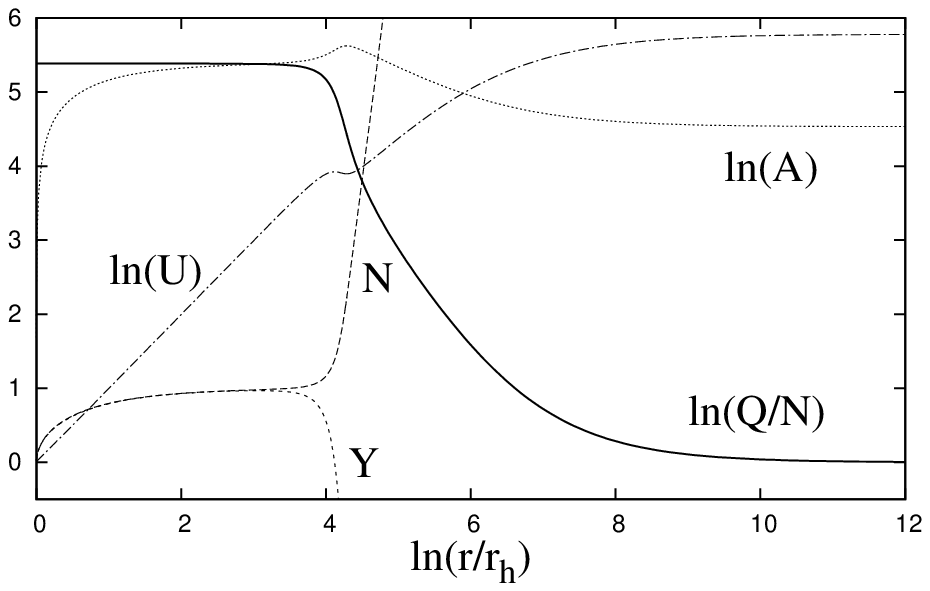}}
	
\hspace{1mm}
\hss}
\caption{{\protect\small 
Hairy deformations of the SAdS background, where $A_0,N_0,Q_0,Y_0$
correspond to the undeformed solution (left);
and of the S background (right). 
    }}
\label{Fig5}
\end{figure}
This is confirmed for the SAdS solutions ($\Lambda_g<0$),
which indeed support a short massive hair and show 
deviations from the pure SAdS 
in the horizon vicinity, but far away from the   
horizon  the deviations tend to zero  (see Fig.\ref{Fig5}).

However, the argument does not work for $\Lambda_g\geq 0$. 
When one deforms the S background by setting $u=r_h+\delta u$
for however small $\delta u$, 
the solutions first stay very close to 
Schwarzschild (until $\ln(r/r_h)\sim 4$ in Fig.\ref{Fig5}). However,
for large $r$  they deviate from the background 
and show a completely different asymptotic behavior at infinity,
characterized by a quasi-AdS g-metric and a compact f-metric giving a finite
value for the volume of the 3-space \cite{Volkov:2012wp}. 
Therefore, the only asymptotically flat black hole in the bigravity theory 
is pure Schwarzschild, and deforming it results in loosing the asymptotic flatness.  
Similarly, trying to deform the SdS background produces 
a curvature singularity at a finite
proper distance away from the black hole horizon,
hence the only asymptotically de Sitter black hole is the pure SdS. 

In the shrinking horizon limit, 
$r_h\to 0$, the black hole `hair' does not disappear but becomes 
a static `lump' made of massive field modes. Such lumps
are described by globally
regular solutions for which the event horizon is replaced by the regular center 
at $r=0$, while at infinity the asymptotic behavior is the same as for the 
black holes \cite{Volkov:2012wp}.  None of the lumps are asymptotically flat
(apart from the flat space). 

\begin{figure}[th]
\hbox to \linewidth{ \hss

	\resizebox{8cm}{5cm}{\includegraphics{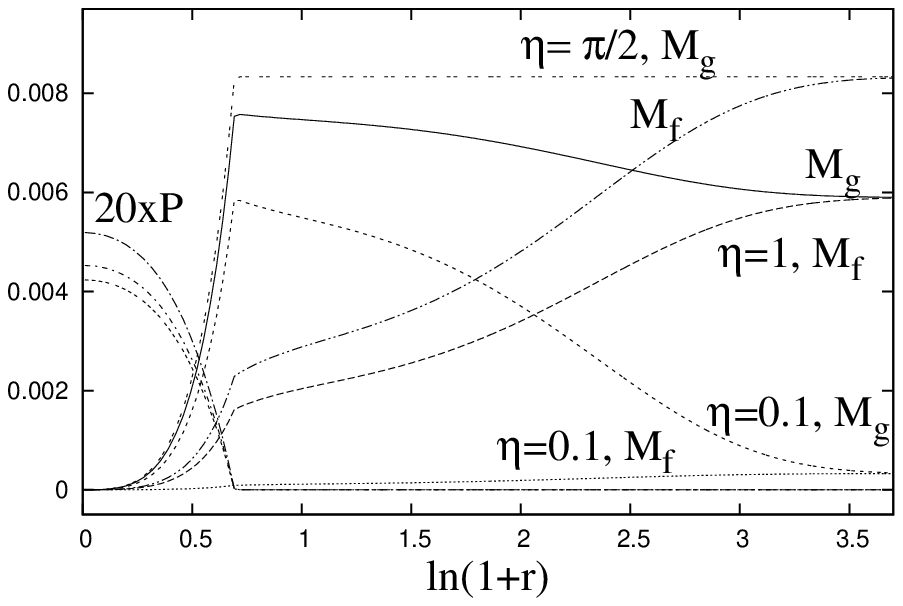}}
\hspace{1mm}
	\psfrag{y}{$U^\prime$}
	\resizebox{8cm}{5cm}{\includegraphics{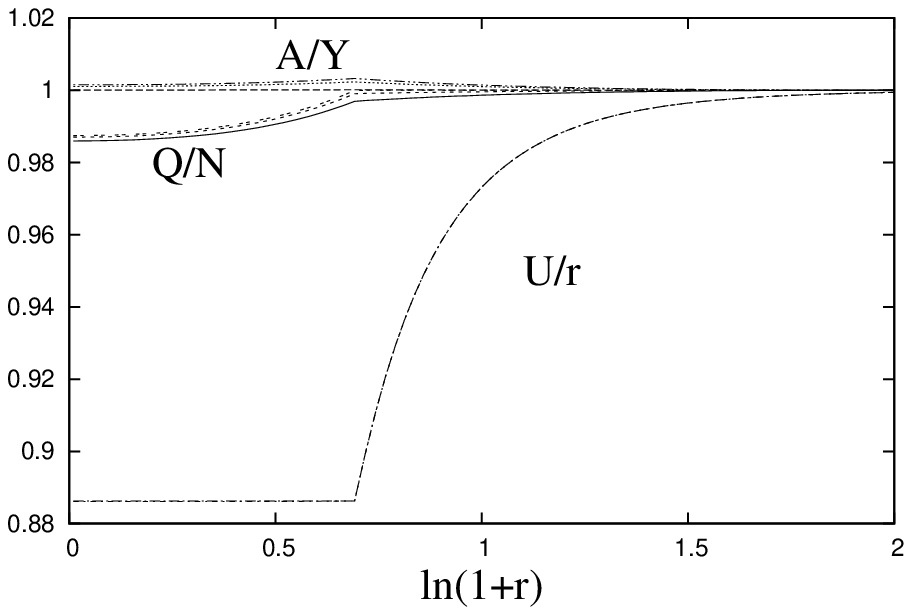}}
	
\hspace{1mm}
\hss}
\caption{{\protect\small Profiles of the  
asymptotically flat star solution sourced by a regular   
matter distribution. 
    }}
\label{Fig6}
\end{figure}

Asymptotically flat solutions can be obtained by adding  matter. Suppose that
the f-sector is empty, while the g-sector contains 
$T^{[{\rm m}]\mu}_{~~~~~\nu}={\rm diag}[-\rho(r),P(r),P(r),P(r)]$ with 
$\rho=\rho_\star\theta(r_\star-r)$, corresponding to 
a `star' with a constant density $\rho_\star$ and a radius $r_\star$.  
Adding this source to the field equations (\ref{eqs}), 
assuming the regular center  at $r=0$ and integrating towards large $r$, 
one finds a solution for which both metrics approach Minkowski metric at infinity. 
Introducing the mass functions $M_g,M_f$ via 
$g^{rr}=N^2=1-2M_g(r)/r$ and  $f^{rr}=Y^2/U^{\prime 2}=1-2M_f(r)/r$, 
one finds that $M_g,M_f$ rapidly increase inside the star, while outside
they approach the same asymptotic value $M_g(\infty)=M_f(\infty)\sim\sin^2\eta$
(see Fig.\ref{Fig6}). 
For $\eta=\pi/2$ the g-metric is coupled only to the matter and is described
by the Schwarzschild solution, $M_g(r)=\rho_\star r^3/6$ for $r<r_\star$ and 
$M_g(r)=\rho_\star r_\star^3/6\equiv M_{\rm ADM}$ for $r>r_\star$. 
For $\eta<\pi/2$ the star mass $M_{\rm ADM}$ is partially screened by the negative 
graviton energy. For $\eta=0$ the f-metric becomes flat, so that $M_f=0$,
while $M_g$ asymptotically approaches zero and the star mass
is totally screened, because the massless graviton decouples and there could be
no $1/r$ terms in the metric. 

If the graviton mass is very small, then the $m^2T^\mu_{~\nu}$ contribution 
to the equations is small as compared to $T^{[{\rm m}]\mu}_{~~~\nu}$, and
for this reason   
$M_g$ rests approximately constant for
$r_\star<r<r_{\rm V}\sim (M_{\rm ADM}/m^2)^{1/3}$. This illustrates the 
Vainshtein mechanism of recovery of General Relativity in a finite region
\cite{Vainshtein:1972sx}. This mechanism has also been independently confirmed 
within the generic massive gravity theory with the BD ghost 
\cite{Babichev:2009jt,Babichev:2010jd},
and in the dRGT theory \cite{Gruzinov:2011mm}.  
It is worth noting that the mechanism works only in the presence of a
regular matter source, and not for vacuum black holes or lumps.    

To recapitulate, globally regular and asymptotically flat stars 
exist both in the bigravity theory and in the dRGT massive gravity.
In the latter case there are also SdS black holes, 
but there are no asymptotically flat black holes. 
There are hairy black holes  in the bigravity theory,
but the only one which is asymptotically flat is the standard ``hairless"
Schwarzschild black hole.

\section{Concluding remarks}

Summarizing what was said above, all known cosmologies and black holes
in the dRGT massive gravity theory are obtained within the  
inhomogeneous St\"uckelberg field approach, assuming non-bidiagonal metrics
\cite{Gumrukcuoglu:2011ew}. Perturbations of the  cosmological solutions  
were studied in \cite{Gumrukcuoglu:2011zh,D'Amico:2012pi,Chiang:2012vh,Wyman:2012iw}.  
 One could perhaps also mention a peculiar FLRW solution \cite{Chamseddine:2011bu}
with a {\it degenerate} f-metric, although its
g-metric is regular and is sourced by the effective 
$T^0_0=\sum _{n=0}^3a_n\Q^{-n}$ with constant $a_n$
(see also \cite{Parisi1,Parisi:2012cg}).

Although massive gravity 
solutions with bidiagonal metrics similar to (\ref{gf1}) 
are very special and hence are 
unlikely to be physically important, 
they are popular due to their simplicity. They have been studied   also 
in the modified versions of the theory, as for example in 
the extended massive gravity 
\cite{Saridakis,Saridakis1,Hinterbichler:2013dv} 
where the graviton mass $m$ is promoted to a dynamical field \cite{D'Amico:2011jj,Huang},
or in the massive gravity  theory with a (quasi) dilaton field \cite{D'Amico:2012zv,Haghani:2013eya}. 
Choosing the fixed f-metric to be non-flat, as for example de Sitter or FLRW, 
also leads to
new cosmological  \cite{Langlois:2012hk,Gong,Motohashi:2012jd,Gumrukcuoglu:2012aa,DeFelice:2013awa,Sakakihara:2012iq} 
and black holes solutions \cite{Gabadadze:2012xv}, and 
allows one 
to study \cite{deRham:2012kf,Fasiello:2012rw,deRham:2013wv} the Higuchi bound 
\cite{Higuchi:1986py}, that is 
 the limit where the scalar polarization of the massive graviton decouples. 
In addition, O(4)-symmetric instanton solutions in the 
Euclidean version of the massive gravity 
theories  
have been considered as well \cite{Zhang:2012ap,Park:2012cq}.  

Changing the reference metric from case to case does not 
seem to be natural. 
If one wants to consider different  possibilities for f, 
then it is logical  to make it dynamical,
which leads to the bigravity. The massive gravities with a fixed f
then can be recovered by choosing a source for f, for example
a constant $\rho_f=P_f>0$, and taking the limit $\eta\to 0$. 
For $\eta\neq 0$ both metrics are dynamical, which gives rise to a much richer 
solution structure than in the massive gravity theories.

\ack

It is a pleasure to acknowledge discussions with Thibault Damour and thank him 
for a careful reading of the manuscript and valuable remarks. I  
would like also to thank  the IHES 
for hospitality
during my visit in March 2013 when this article was written, and I 
thank Shinji Mukohyama for inviting me to write it for Classical and Quantum
Gravity.

\section*{References}



\end{document}